\documentclass[useAMS,usenatbib,usegraphicx]{mn2e}

\def\k{km s$^{-1}$}
\def\ks{km s$^{-1}$~}
\def\d{$^\circ$}
\def\m{$^\prime$}
\def\s{$^{\prime\prime}$}
\def\hh{$^{\mathrm h}$}
\def\mm{$^{\mathrm m}$}
\def\ss{$^{\mathrm s}$}
\def\cm3{cm$^{-3}$}

\def\2{$^{12}$CO}
\def\3{$^{13}$CO}
\def\msol{M$_\odot$}

\title[Study of the molecular gas surrounding the EGO G35.03$+$0.35]
  {Study of the dense molecular gas surrounding the ``Extended Green Object'' G35.03$+$0.35}

\author[S. Paron, M. E. Ortega, A. Petriella, M. Rubio, E. Giacani and G. Dubner]{S. Paron$^{1,2}$\thanks{E-mail:
sparon@iafe.uba.ar}, M. E. Ortega$^{1}$, A. Petriella$^{1}$, M. Rubio$^{3}$, E. Giacani$^{1,2}$ and G. Dubner$^{1}$\\
$^{1}$ Instituto de Astronom\'\i a y  F\'\i sica del Espacio
(CONICET-UBA), CC 67, Suc. 28, 1428 Buenos Aires, Argentina\\
$^{2}$ FADU - Universidad de Buenos Aires\\ 
$^{3}$ Departamento de Astronom\'{\i}a, Universidad de Chile, Casilla 36-D, Santiago, Chile\\}  
\begin{document}

\date{Accepted XXXX. Received XXXX; in original form XXXX}

\pagerange{\pageref{firstpage}--\pageref{lastpage}} \pubyear{2011}

\maketitle

\label{firstpage}

\begin{abstract}

We present the results of a new study of the molecular gas associated with the ``extended green object'' (EGO) G35.03$+$0.35. 
This object, very likely a massive young stellar object, is embedded in a molecular cloud at the border of an HII region. 
The observations were performed with the Atacama Submillimeter Telescope Experiment (ASTE) in the $^{12}$CO and $^{13}$CO J=3--2,
HCO$^{+}$ J=4--3, and CS J=7--6 lines with an angular resolution about 22\s. 
From the $^{12}$CO J=3--2 line we discovered outflowing activity of the massive young
stellar object. We obtained a total mass and kinetic energy for the outflows of 30 \msol~and 3000 \msol [\k]$^{2}$ 
($6 \times 10^{46}$ ergs), respectively. We discovered a HCO$^{+}$ and CS clump towards the EGO G35.03$+$0.35. 
The detection of these molecular species supports the presence of molecular outflows and a dense molecular envelope with temperatures and densities 
above 40 K and $6 \times 10^{6}$ cm$^{-3}$, respectively. 
Using public near- and mid-IR, and sub-mm data we investigated the 
spectral energy distribution confirming that EGO G35.03$+$0.35 is a massive young stellar object at the earliest evolutionary stage (i.e. a class I
young stellar object). By anlysing radio continuum archival data we found three radio sources towards the object, suggesting the presence of several 
young stellar objects in the region. Our radio continuum analysis is consistent with the presence of at least one ultracompact HII region and an 
hypercompact HII region or a constant-velocity ionized wind source.

\end{abstract}


\section{Introduction}

\begin{figure*}
\centering
\includegraphics[width=12cm]{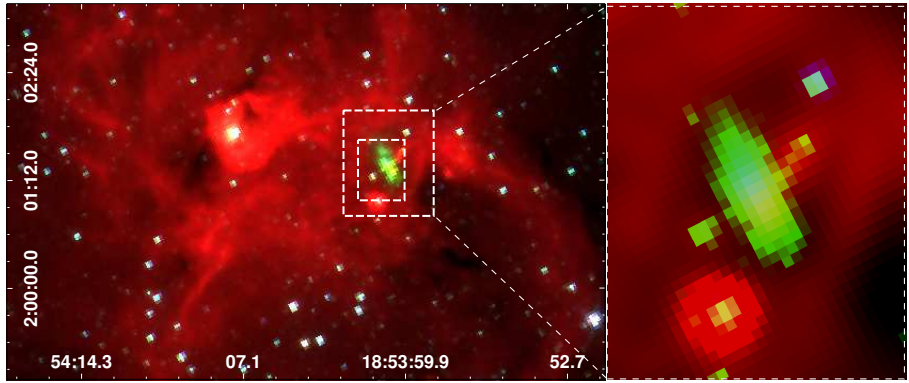}
\caption{Left: {\it Spitzer}-IRAC three color image
(3.6 $\mu$m $=$ blue, 4.5 $\mu$m $=$ green and 8 $\mu$m $=$ red). The dashed rectangles are the regions mapped in \2 J=3--2 and
HCO$^{+}$~J=4--3 (the largest one) and in \3 J=3--2 and CS J=7--6 (the smallest one). Right: zoom in of the surveyed region.  }
\label{present}
\end{figure*}

\citet{cyga08} identified more than 300 Galactic extended 4.5 $\mu$m sources,
naming extended green objects (EGOs) or ``green fuzzies'', for the common coding of the [4.5 $\mu$m] band as green in three-color
composite Infrared Array Camera images from the {\it Spitzer} Telescope.
According to the authors, an EGO is a probable massive young stellar object (MYSO) driving outflows.
The extended emission in the 4.5  $\mu$m band is supposed to be due to H$_{2}$ ($\nu=0-0$, S(9,10,11)) lines and CO ($\nu = 1-0$) band heads,
that are excited by the shock of the outflows propagating in the interstellar medium (see \citealt{noriega04,marston04,smith05}).
Recently, \citet{debuizer10} reported the first direct spectroscopic identification of the origin of the 4.5 $\mu$m emission towards two
EGOs using NIRI on the Gemini North telescope. In one of the observed EGOs, they proved that the 4.5 $\mu$m emission is due primarily to
lines of molecular hydrogen, which are collisionally excited.

EGO G35.03+0.35 (hereafter EGOg35) is embedded in a dense molecular cloud  at the distance of 3.5 kpc \citep{albert10} located on
the border of the infrared dust bubble N65 \citep{church06,church07}. According to \citet{albert10} there are several young stellar object
(YSO) candidates around N65, being EGOg35 the most prominent source. Using IR and sub-mm fluxes measured from this EGO, the authors performed
an spectral energy distribution (SED), showing that this source is indeed a massive stellar object at the earlier stages of evolution with
outflowing activity. The 4.5 $\mu$m emission of this EGO has a bipolar morphology, with one lobe to the NE and the other to the SW. 
The source presents maser emission of several molecular lines (see e.g. \citealt{forster89,caswell95,kurtz05}). 
Recently, CH$_{3}$OH maser emission at 6.7 and 44 GHz has been detected and 
analysed by \citet{cyga09}. The 6.7 GHz CH$_{3}$OH masers are radiatively pumped by IR emission 
from the warm dust associated exclusively with massive YSOs,
while the methanol maser at 44 GHz is collisionally excited in molecular outflows, and particularly at interfaces between outflows
and the surrounding ambient cloud (see \citealt{cyga09} and references therein).

Considering that EGOg35 is indeed a massive YSO evolving in a dense molecular medium, it is important to understand the physical 
characteristics of its ambient medium. We investigated this ambient 
through several molecular lines obtained with the Atacama Submillimeter Telescope Experiment (ASTE). The results of this investigation 
are presented in this work.

\section{Observations}

The molecular observations were performed on July 14 and 15, 2010 with the 10 m Atacama Submillimeter
Telescope Experiment (ASTE; \citealt{ezawa04}). We used the CATS345 GHz band receiver, which is a two-single
band SIS receiver remotely tunable in the LO frequency range of 324-372 GHz. We simultaneously
observed \2 J=3--2 at 345.796 GHz and HCO$^{+}$~J=4--3 at
356.734 GHz, mapping a region of 70\s~$\times$ 80\s~centered at RA $=$ 18\hh 54\mm 0.7\ss, dec. $= +$02\d 01\m 18.9\s, J2000).
We also observed \3 J=3--2 at 330.588 GHz and CS J=7--6 at 342.883 GHz towards the same center mapping a region
of 40\s~$\times$ 50\s. The mapping grid spacing was 10\s~and the integration time was 60 sec per pointing in both cases.
All the observations were performed in position switching mode. We verified that the off position (RA $=$  18\hh 53\mm 55.8\ss,
dec. $= +$02\d 07\m 49.5\s, J2000) was to be free of emission.

We used the XF digital spectrometer with a bandwidth and spectral resolution set to 128 MHz and 125 kHz, respectively.
The velocity resolution was 0.11 \ks and the half-power beamwidth (HPBW) was 22\s, for all observed molecular lines. The system temperature
varied from T$_{\rm sys} = 150$ to 200 K. The main beam efficiency was $\eta_{\rm mb} \sim 0.65$.
The spectra were Hanning smoothed to improve the signal-to-noise ratio and only linear or/and some third order
polynomia were used for baseline fitting.
The data were reduced with NEWSTAR and the spectra processed using the XSpec software package.

To complement the molecular data we use near- and mid-IR and radio continuum data from public databases and catalogues, which 
are described in the corresponding sections.

\section{Results and discussion}

The source EGO G35.03$+$0.35 (EGOg35) is located at the border of an HII region which is delineated mainly by the 8 $\mu$m emission.
Figure \ref{present} (left) shows a composite three-color image of a region towards EGOg35. The image displays
three {\it Spitzer}-IRAC bands: 3.6 $\mu$m (in blue), 4.5 $\mu$m (in green) and 8 $\mu$m (in red). EGOg35 is the green structure
inside the dashed rectangles, which represent the 70\s~$\times$ 80\s~and 40\s~$\times$ 50\s~regions mapped in the molecular
lines as described in the previous section. A zoom in of the surveyed region is shown in Fig. \ref{present} (right).

\subsection{Molecular lines}
\label{moleclines}

\begin{figure}
\centering
\includegraphics[width=8.3cm]{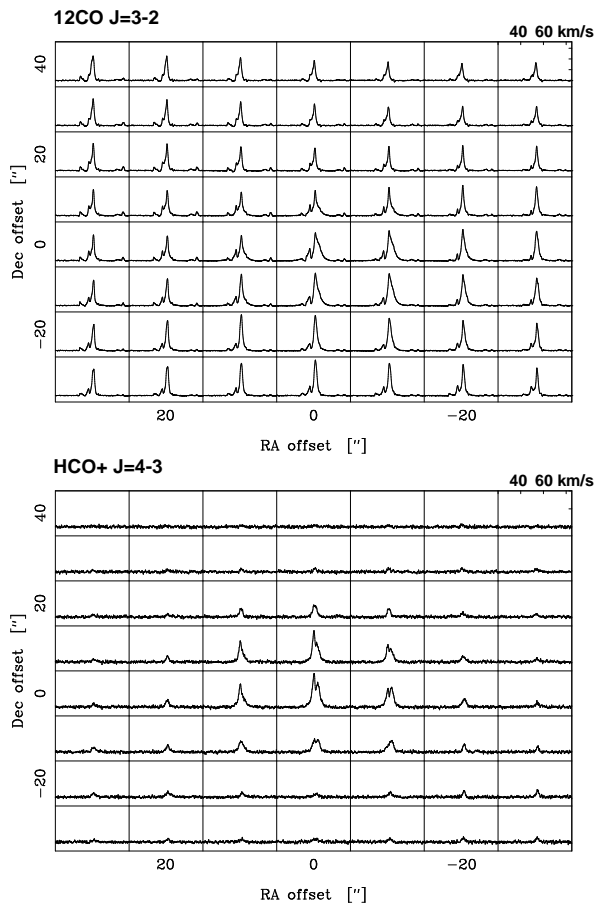}
\caption{\2 J=3--2 and HCO$^{+}$ J=4--3 spectra obtained towards EGOg35.}
\label{obs1}
\end{figure}

\begin{figure}
\centering
\includegraphics[width=8cm]{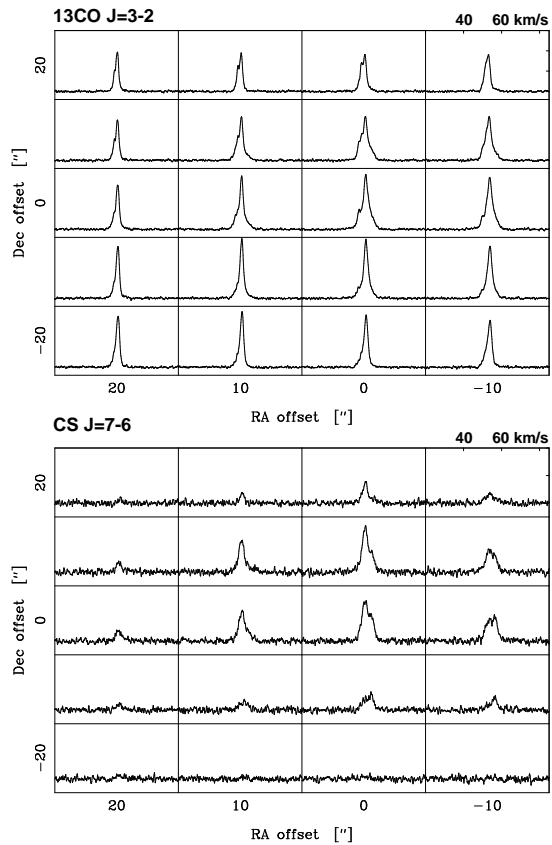}
\caption{\3 J=3--2 and CS J=7--6 spectra obtained towards EGOg35.}
\label{obs2}
\end{figure}

To study the molecular ambient where EGOg35 is embedded, we analyse the \2 and \3 J=3--2, 
HCO$^{+}$~J=4--3 and CS J=7--6 transitions, tracers of outflows and dense gas.
Figures \ref{obs1} and \ref{obs2} display the molecular lines spectra obtained towards EGOg35. Most of the spectra are far of having 
a simple Gaussian shape, presenting asymmetries, 
probable absorption dips, and spectral wings or shoulders, which suggest that the 
molecular gas is affected by the dynamics of EGOg35. In what follows we study the gas kinematics.

\begin{figure}
\centering
\includegraphics[width=8.5cm]{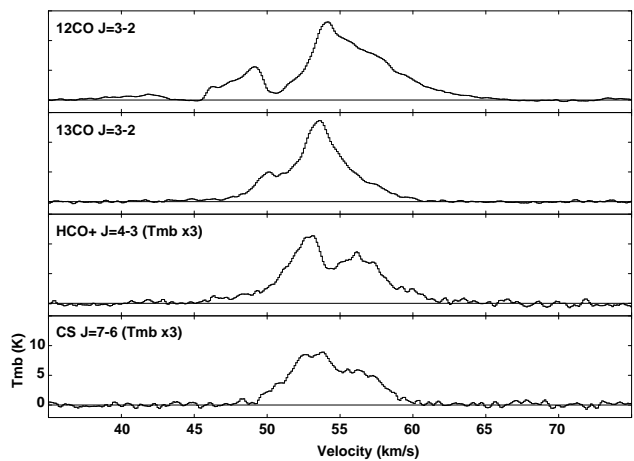}
\caption{Spectra towards the center of the observed region. The
HCO$^{+}$ J=4--3 and CS J=7--6 lines were scaled with a factor of $\times3$. The vertical axis is main beam brightness temperature.}
\label{spec00}
\end{figure}

\begin{figure}
\centering
\includegraphics[width=7cm]{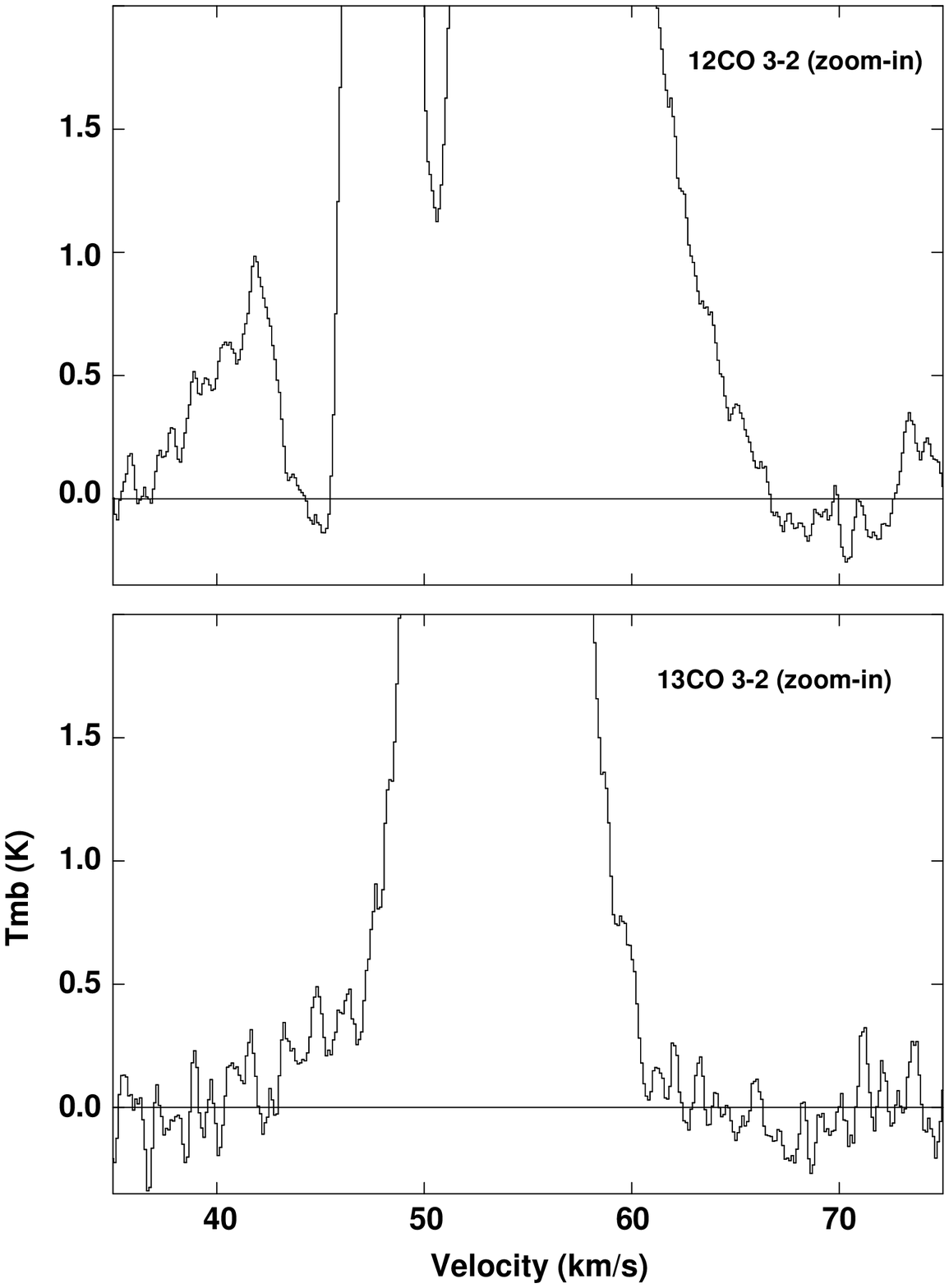}
\caption{Zoom-in of the \2 and \3 J=3--2 profiles presented in Fig. \ref{spec00} in the intensity range going 
from $-0.35$ to 2 K.}
\label{spec00zoom}
\end{figure}

Figure \ref{spec00} shows the spectra obtained towards the central position of EGOg35, that is the (0,0) offset in Figs. \ref{obs1} and \ref{obs2}.
In the case of the HCO$^{+}$ J=4--3 and CS J=7--6 lines, their brightness temperatures were scaled with a factor of $\times3$. 
Figure \ref{spec00zoom} displays a zoom-in of the central \2 and \3 J=3--2 profiles in the intensity range that goes 
from $-0.35$ to 2 K, in order to note some weak features. In the case of the \2 line can be appreciated a weak component ($\sim5$ times above 
the rms noise) centered at $\sim41$ \k. 
The parameters determined from Gaussian fitting of these lines are presented in Table \ref{lines}.
T$_{mb}$ represents the peak brightness temperature, V$_{LSR}$ the central velocity referred to the Local Standard of Rest. 
Errors are formal 1$\sigma$ value for the model of the Gaussian line shape. All the lines
were well fit with more than one Gaussian function, which very likely indicates the presence of several components or/and 
spectral wings, usually signatures of outflows. In Table \ref{lines2} we present V$_{max}$ and V$_{min}$, 
indicating the respective total widths of the spectra considering all the components, and $\int{T_{mb} dv}$, the intensity integrated over
the whole profile.

The \2 J=3--2 spectrum presented in Fig. \ref{spec00} shows a double peak structure like with a main component centered at $\sim$54.2 \ks 
and a less intense component centered at $\sim$48.7 \k. The \3 J=3--2 spectrum  presents similar features. We conclude that these lines
present an absorption dip at $\sim$51.5 \ks which separates both mentioned velocity components, showing that the lines are self-absorbed as it is 
usually found towards star-forming regions \citep{johnstone03,buckle10,ortega10}. 
The velocity of the mentioned \2 and \3 dip is coincident (within the errors) with the central velocities of the SiO (5--4), H$^{13}$CO$^{+}$ (3--2), 
and CH$_{3}$OH (5$_{2,3}$--4$_{1,3}$) lines observed towards the center 
of EGOg35 by \citet{cyga09} and with the CS J=7--6 main component reported in this work. Therefore, we conclude that v $\sim$51.5 \ks is the velocity 
of the ambient gas, and the other reported components (see Table \ref{lines}) may be related to outflows or/and high velocity material. 
In Section \ref{secoutflow} we analyse this contention. 

\begin{table}
\caption{Observed parameters of the molecular lines shown in Figure \ref{spec00}.}
\begin{minipage}{140mm}
\begin{tabular}{lcc}
\hline
Emission & T$_{mb}$ & V$_{LSR}$    \\
         &  (K)     & (\k)        \\
\hline
$^{12}$CO J=3--2 & 0.80 $\pm$0.20   &  41.12 $\pm$0.56  \\
                 & 4.55 $\pm$0.25   &  48.68 $\pm$1.25  \\
                 & 12.40 $\pm$0.75  &  54.20 $\pm$1.20  \\
                 & 4.42 $\pm$1.50   &  56.90 $\pm$1.20  \\
                 & 3.65 $\pm$0.85   &  59.10 $\pm$1.25  \\
\hline
$^{13}$CO J=3--2  & 4.40 $\pm$0.45  &  50.10 $\pm$1.60   \\
                & 12.85 $\pm$1.70  &  53.50 $\pm$1.40    \\
                & 2.48 $\pm$0.60  & 56.75 $\pm$1.50   \\
\hline
HCO$^{+}$ J=4--3  & 3.10 $\pm$0.8 & 52.55 $\pm$0.75  \\
                  & 2.60 $\pm$0.8 & 56.05 $\pm$0.85  \\
\hline
CS J=7--6       & 2.85 $\pm$0.35  & 53.15 $\pm$0.65  \\
                & 1.45 $\pm$0.60  & 57.05 $\pm$0.50  \\
\hline
\end{tabular}
\label{lines}
\end{minipage}
\end{table}

\begin{table}
\caption{Derived parameters of the molecular lines shown in Figure \ref{spec00}.}
\begin{minipage}{140mm}
\begin{tabular}{lccc}
\hline

Emission         &  V$_{min}$ & V$_{max}$ & $\int{T_{mb}dv}$ \\
                 &   (\k)     &   (\k)    &  (K \k)          \\
\hline
$^{12}$CO J=3--2 &    37.0    &  67.0     &   92.4 $\pm$1.9   \\
$^{13}$CO J=3--2 &    44.5    &  60.2     &   64.2 $\pm$1.8   \\
HCO$^{+}$ J=4--3 &    45.5    &  61.6     &   15.5 $\pm$0.7  \\
CS J=7--6        &    47.7    &  60.7     &   16.6 $\pm$0.8  \\
\hline
\end{tabular}
\label{lines2}
\end{minipage}
\end{table}

\begin{figure}
\centering
\includegraphics[width=7cm,angle=-90]{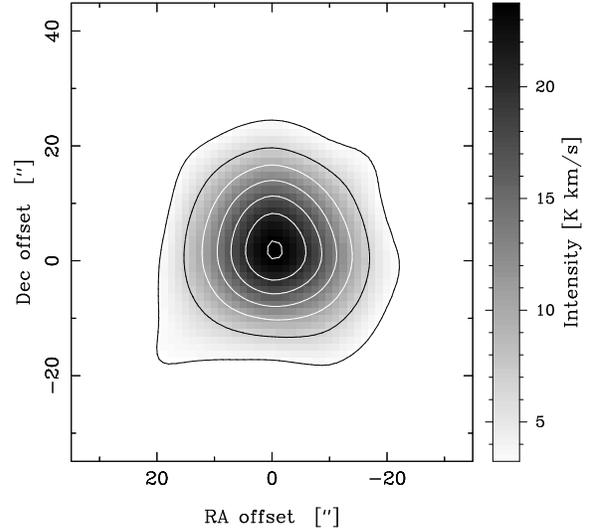}
\caption{HCO$^{+}$ J=4--3 emission integrated between 45 and 62 \k. The beam size is 22\s. }
\label{hco+}
\end{figure}

Regarding the HCO$^{+}$ J=4--3 emission, the profiles towards the center of the surveyed region  
present a dip at v $\sim54$ \ks (see Fig. \ref{obs1} bottom, and Fig. \ref{spec00}).  
\citet{cyga09} presented a single point observation of H$^{13}$CO$^{+}$ J=4--3 towards EGOg35. They report
that the H$^{13}$CO$^{+}$ J=4--3 emission peaks at $\sim53.1$ \ks with a FWHM $\Delta$v $\sim5.4$ \k. Taking into account
that the H$^{13}$CO$^{+}$ emission is optical thinner than the HCO$^{+}$ emission and peaks at a velocity close to 
that of the HCO$^{+}$ dip, we conclude that this emission indeed appears self-absorbed. This kind of spectral feature 
might be revealing the existence of a density gradient in the clump \citep{hira07}, consistent with the presence
of an embedded central source with outflowing activity. It is known that such molecular species enhances
in molecular outflows \citep{raw04}. In effect, a strong enhancement of the HCO$^{+}$ abundance is
expected to occur in the boundary layer between the outflow jet and the surrounding
molecular core. This would be due to the liberation and photoprocessing by the shock of the molecular material
stored in the icy mantles of the dust. Figure \ref{hco+} displays the HCO$^{+}$ J=4--3 emission integrated between 
45 and 62 \k, showing a HCO$^{+}$ molecular clump peaking at the position of EGOg35.

\begin{figure}
\centering
\includegraphics[width=7cm,angle=-90]{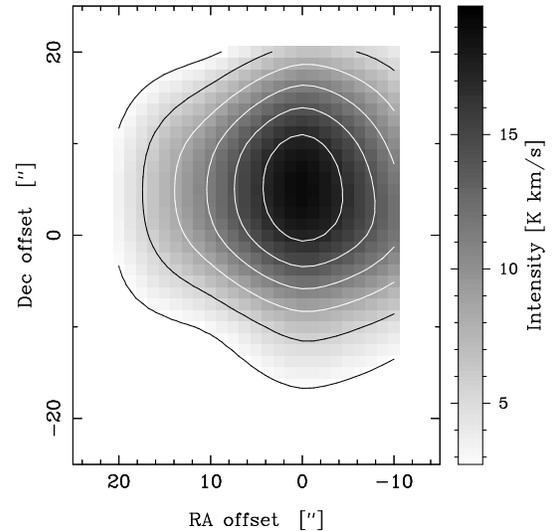}
\caption{CS J=7--6 emission integrated between 45 and 62 \k. The beam size is  22\s.  }
\label{cs}
\end{figure}

It is worth noting that the central HCO$^{+}$ and CS spectra, which survey the densest region where EGOg35 is embedded, have 
the closer component (i.e. the component with lower velocity, or blue component) stronger than the farthest one 
(red component). This signature suggests the presence of infalling gas, which is consistent with the presence of a YSO accreting material. 
As \citet{zhou93} explain, this effect reflects the fact that the excitation of molecules is higher in the cloud center and, if the 
cloud has infalling 
motion, then the observer is looking at the hotter side of the blue hemisphere and the cooler side of the red hemisphere. Therefore, the blue emission 
should always be as strong as, or stronger than, the red emission.

Finally, it is important to note that the CS J=7--6 emission maps the dense envelope where the massive YSO is evolving. The detection of this line 
implies the presence of a gaseous envelope with temperatures and densities above 40 K and $6 \times 10^{6}$ cm$^{-3}$, respectively 
(e.g. \citealt{taka07}). Figure \ref{cs} shows the CS J=7--6 emission integrated between 45 and 62 \k.

\subsubsection{Column densities and abundances}

To estimate the molecular column densities and hence the abundances in the region we assume as a first order approximation local
thermodynamic equilibrium (LTE) and a beam filling factor of 1.
From the peak temperature ratio between the CO isotopes ($^{12}$T$_{mb}$/$^{13}$T$_{mb}$; taken from the CO main components at v $\sim54$ \k), 
it is possible to estimate the optical depths from (e.g. \citealt{scoville86}):
\begin{equation}
\frac{^{12}{\rm T}_{mb}}{^{13}{\rm T}_{mb}} = \frac{1-exp(-\tau_{12})}{1-exp(-\tau_{12}/X)}
\label{eq1}
\end{equation}
where $\tau_{12}$ is the optical depth of the \2 gas and $X =$ [\2]/[\3] is the isotope abundance ratio. Assuming R$_{\odot} = 8$ and
using [\2]/[\3] $= (6.21\pm1.00){\rm D_{GC}} + (18.71\pm7.37)$  \citep{milam05} where D$_{\rm GC} = 5.5$ kpc is the distance between
the source and the Galactic Center, we obtain [\2]/[\3] $ = 53 \pm 12$. Then the \2 J=3--2 optical depth
is $\tau_{12} \sim 300$, while the \3 J=3--2 optical depth is $\tau_{13} \sim 6$, revealing that both lines are optically thick.
Thus, we calculate the excitation temperature from
\begin{equation}
T_{ex}(3 \rightarrow  2) = \frac{16.95 {\rm K}}{{\rm ln}[1 + 16.59 {\rm K} / (T_{\rm max}(^{12}{\rm CO}) + 0.036 {\rm K})]}
\label{eq2}
\end{equation}
obtaining $T_{ex} \sim 20$ K. Then, we derive de \3 and \2 column densities from (see e.g. \citealt{buckle10}):
\begin{equation}
{\rm N(^{13}CO)} = 8.28 \times 10^{13}~e^{\frac{15.87}{T_{ex}}}\frac{T_{ex} + 0.88}{1 - exp(\frac{-15.87}{T_{ex}})} \int{\tau_{13}{\rm dv}}
\label{eq3}
\end{equation}
and 
\begin{equation}
{\rm N(^{12}CO)} = 7.96 \times 10^{13}~e^{\frac{16.6}{T_{ex}}}\frac{T_{ex} + 0.92}{1 - exp(\frac{-16.6}{T_{ex}})} \int{\tau_{12}{\rm dv}}
\label{eq4}
\end{equation}
where, taking into account that $\tau \geq 1$ in both cases, we use the approximation:
\begin{equation}
\int{\tau ~{\rm dv}} = \frac{1}{J(T_{ex}) - J(T_{\rm BG})} \frac{\tau}{1-e^{-\tau}} \int{{\rm T_{mb} ~dv}}
\label{eq5}
\end{equation}
with 
\begin{equation}
J(T) = \frac{h\nu/k}{exp(\frac{h\nu}{kT}) - 1}
\label{eq6}
\end{equation}
We obtain N(\3) $\sim 7 \times 10^{16}$ cm$^{-2}$ and  N(\2) $\sim 1 \times 10^{19}$ cm$^{-2}$.

Additionally, we derive the CS J=7--6 and HCO$^{+}$ J=4--3 column densities from: 
\begin{equation} 
{\rm N(CS)} = 2.54 \times 10^{11}~e^{\frac{49.4}{T_{ex}}}\frac{T_{ex} + 0.56}{1 - exp(\frac{-16.45}{T_{ex}})} \int{\tau{\rm dv}}
\label{eq7}
\end{equation}
and
\begin{equation}
{\rm N(HCO^{+})} = 5.85 \times 10^{10}~e^{\frac{25.7}{T_{ex}}}\frac{T_{ex} + 0.71}{1 - exp(\frac{-17.12}{T_{ex}})} \int{\tau{\rm dv}}
\label{eq8}
\end{equation}
Since the CS J=7--6 line is optically thick (e.g. \citealt{giannini05}), we use again the approximation presented in 
equation (\ref{eq5}) assuming $\tau_{\rm CS} = 1$. On the other hand, by assuming that the HCO$^{+}$ J=4--3 is optically thin, we use the approximation:
\begin{equation}
\int{\tau ~{\rm dv}} = \frac{1}{J(T_{ex}) - J(T_{\rm BG})} \int{{\rm T_{mb} ~dv}}
\label{eq9}
\end{equation}
We adopt as excitation temperatures 66 K and 43 K for the CS and HCO$^{+}$, respectively, which correspond to the equivalent 
temperature of each molecular transition. We therefore obtain N(CS) $ \sim 8 \times 10^{13}$ cm$^{-2}$ 
and N(HCO$^{+}$) $\sim 9.5 \times 10^{12}$ cm$^{-2}$.

Assuming the abundance ratio [H$_{2}$]/[\3] $= 77 \times 10^{4}$ \citep{wilson94}, from the N(\3) we can estimate the
H$_2$ column density in N(H$_{2}$) $\sim 5 \times 10^{22}$ cm$^{-2}$. This value is in agreement with the  H$_{2}$ column densities 
reported for high-mass protostar candidates associated with methanol masers \citep{codella04,sz07}, 
as in our case. Thus, using this H$_{2}$ column density we derive the abundance ratios for the HCO$^{+}$ and CS:
X(HCO$^{+}$) $\sim 2 \times 10^{-10}$ and X(CS) $\sim 1.6 \times 10^{-9}$. The X(CS) value is within the very wide range ($5 \times 10^{-10} - 
2 \times 10^{-7}$) measured in the outflows of protostars (e.g. \citealt{botti04,jorgensen04,giannini05}).

\subsubsection{High velocity material and outflows}

\begin{figure}
\centering
\includegraphics[width=7cm,angle=-90]{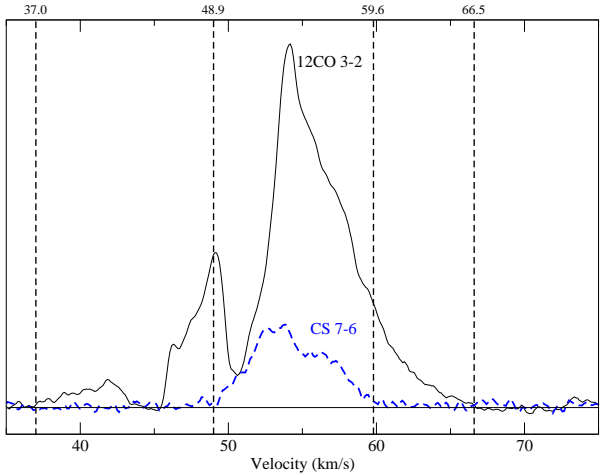}
\caption{\2 J=3--2 (solid line) and CS J=7--6 (dashed line) towards the center of EGOg35.  The vertical lines remark the ranges 
where \2 emission is detected at higher and lower velocities with respect to the CS.}
\label{cocs}
\end{figure}

As discussed in Section \ref{moleclines}, the \2 J=3--2  spectrum obtained towards the center of EGOg35 
shows several components, some of them suggesting outflowing activity from EGOg35. The presence of outflows or high velocity material moving 
along the line of sight can be proven by comparing 
the \2 emission with the higher density tracer CS J=7--6. Figure \ref{cocs} presents the \2 and CS spectra towards the center of the 
region with vertical lines remarking the ranges where \2 emission is detected at higher and lower velocities with respect to the CS. 
This figure confirms the presence of \2 spectral wings, which should be due to a red outflow going from 59.6 to 66.5 \ks and a blue one, extending
from 37.0 to 48.9 \k. In the case of the blue wing, we consider that the weak \2 component centered at $\sim$41 \ks is part of the source outflows 
because its velocity is compatible with that of the 6.7 GHz methanol maser detected by \citet{cyga09}, confirming that this \2 component 
is due to the gas expansion caused by EGOg35.

\begin{figure}
\centering
\includegraphics[width=6.5cm]{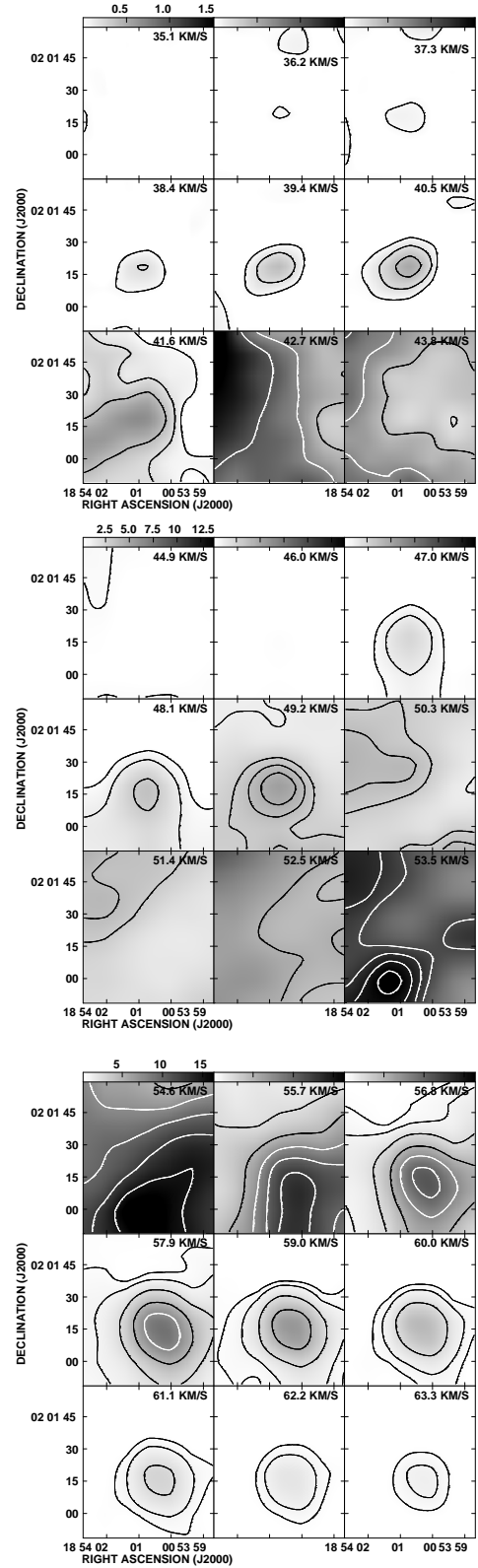}
\includegraphics[width=6.5cm]{12coPaneles1.ps}
\includegraphics[width=6.5cm]{12coPaneles2.ps}
\caption{Integrated velocity channel maps of the \2 J=3--2 emission every $\sim$ 1.1 \k.
The grayscale is displayed at the top of each figure and is in K \k, the contour levels are 0.05, 0.2, 0.4, and 0.7 K \ks for the first panel, 
0.5, 1, 3, 4, 4.5, 10, 11, 12.5 K \ks for the second panel, and 0.5, 1, 2.5, 5, 7.5, 10, 12.5, 15 K \k, for the third one. The beam size is  22\s.}
\label{panels}
\end{figure}

\begin{figure}
\centering
\includegraphics[width=6.5cm]{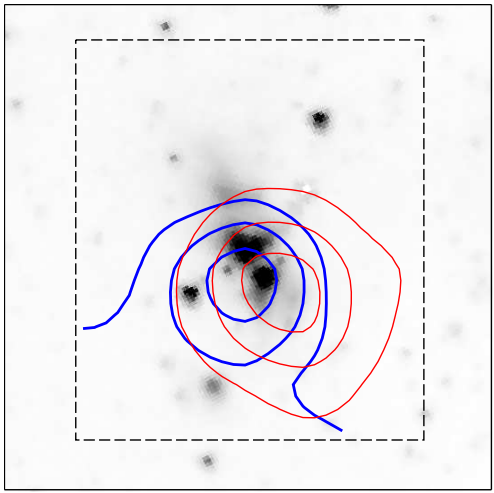}
\caption{In grays is displayed the 4.5 $\mu$m emission from EGOg35. The blue thick and the red thin contours show the \2 J=3--2 emission 
integrated from 37 to 49 \ks and 58 to 66 \k, respectively. The levels are 6.5, 9.5, and 12.5 K \ks for the blue emission and 
4, 10, and 16 K \ks for the red one. The dashed square is the surveyed area. }
\label{outflows}
\end{figure}

To analyze the velocity and spatial distribution of the \2 J=3--2 emission, in Fig. \ref{panels} we present integrated velocity
channel maps every $\sim$1.1 \k. The spatial distributions of the blue and red wings are shown in channels going from $\sim36$ \ks to 50 \ks, 
and $\sim57$ \ks to 64 \k, respectively. Figure \ref{outflows} displays the 4.5 $\mu$m emission with contours of the \2 J=3--2 
line integrated from 37 to 49 \ks and 58 to 66 \ks (blue thick and red thin contours, respectively). 

In what follows, from the \2 spectral wings we estimate the mass and energy to study the dynamics involved in the EGOg35 outflowing 
activity. Using  $M = \mu m_{H} d^{2} \Omega X({\rm CO})^{-1} {\rm N(CO)}$, we obtain the mass for 
the red and blue molecular outflows. N(CO) is the \2 column density, $d$ the distance,
$m_{H}$ the hydrogen atom mass, we adopt a mean molecular weight per H$_{2}$ molecule of $\mu = 2.72$ to include helium, $\Omega$ is the 
area of the \2 red and blue clumps shown in Figs. \ref{panels} and \ref{outflows}, and $X({\rm CO}) = 10^{-4}$ is the \2 relative 
abundance \citep{frerking82}. Integrating the \2 emission from 58 to 66 \ks we obtain a N(CO)$_{red} \sim 2.1 \times 10^{17}$ cm$^{-2}$ 
and $M_{red} \sim 5$ \msol, while integrating from 37 to 49 \ks we obtain N(CO)$_{blue}$ $\sim 1.1 \times 10^{18}$ cm$^{-2}$
and $M_{blue} \sim 24$ \msol.
We calculate the momentum and energy of the red and blue components using: 
\begin{equation}
P = M V
\label{eq10}
\end{equation}
\begin{equation}
E_{k} = \frac{1}{2} M V^{2}
\label{eq11}
\end{equation}
where $V$ is a characteristic velocity estimated as the difference between the maximum velocity of detectable \2 emission in the red and 
blue wings respectively, and the molecular ambient velocity ($\sim51.5$ \k), being $V_{red} \sim 14.5$ \ks and
$V_{blue} \sim 14.5$ \k. Thus, we obtain $P_{red} = 72$ \msol \ks and $E_{k}^{red} = 510$ \msol [\k]$^{2}$ ($E_{k}^{red} \sim 1 \times 10^{46}$ 
ergs), and $P_{blue} = 350$ \msol \ks and $E_{k}^{blue} = 2500$ \msol [\k]$^{2}$ ($E_{k}^{blue} \sim 5 \times 10^{46}$ ergs). The obtained outflows 
parameters are summarized in Table \ref{outf}.

\begin{table}
\caption{Outflow parameters.}
\begin{minipage}{120mm}
\begin{tabular}{lcccc}
\hline

Shift         &  N(CO)                      &   M      &  $P$       &   $E_{k}$             \\
              &  ($\times 10^{18}$ cm$^{-2}$) &  (\msol)   &  (\msol \k)  &   (\msol [\k]$^{2}$)    \\
\hline
Red           &  0.21                        &  5       &  72        &   510  \\
Blue          &  1.10                        &  24      &  350       &   2500 \\
\hline
\end{tabular}
\label{outf}
\end{minipage}
\end{table}

The derived mass and energy for the outflows discovered towards EGOg35 are similar
to those of massive and energetic molecular outflows driven by high-mass YSOs \citep{beuther02,wu04}.
We do not provide estimates of outflow dynamical timescale and mass rate because the analysed high velocity 
material is aligned along the line of sight, and therefore it is not possible to estimate the lengths of the outflow lobes.
In fact, with the moderate angular resolution of the present observations, it is possible that we miss the emission of the outflows along 
the plane of the sky. Molecular observations with higher angular resolution are needed to spatially resolve the outflows.

\subsection{Spectral energy distribution}
\label{secoutflow}

\begin{figure}
\centering
\includegraphics[width=6.5cm]{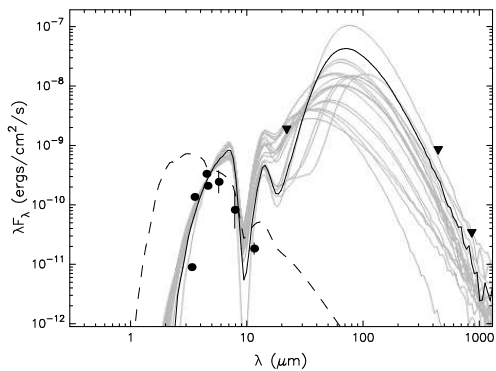}
\caption{Spectral energy distributions of the EGO G35.03+0.35. The black line correspond to the best fitting model and the
gray lines to the following 19 best models. The dashed line is the photosphere model of the central
source in the best fitting model. }
\label{sed1}
\end{figure}

\citet{albert10} performed a spectral energy distribution (SED) of EGOg35 concluding that the source is indeed a massive YSO.
Taking into account that at present it is available new mid-IR fluxes obtained from WISE Preliminary Release Source 
Catalog\footnote{The Wide-field Infrared Survey 
Explorer (WISE) is a joint project of the University of California, Los Angeles, and the Jet Propulsion
Laboratory/California Institute of Technology, funded by the National Aeronautics and Space Administration.}, in this section we 
perform a new SED of the source to study with more accuracy its physical parameters. The advantage of this new SED is that the WISE
data present fluxes at 3.4, 4.6, 12, and 22 $\mu$m bands, being of importance the last two bands, mainly the 12 $\mu$m, because 
the SED of a YSO usually presents the separation between the contributions from the disk and 
envelope fluxes around this wavelength.
Thus, we fit the SED using the tool developed by \citet{robi07}\footnote{http://caravan.astro.wisc.edu/protostars/}. 
We use the fluxes in the four {\it Spitzer}-IRAC bands obtained from \citet{cyga08} together with the WISE fluxes,
and the SCUBA bands at 450 and 850 $\mu$m (source G35.02+0.35, \citealt{hill06}).
The fluxes at 22 $\mu$m from WISE and SCUBA bands were taken as upper limits as they have lower angular resolution
and they can include contributions from other sources around EGOg35. As done in \citet{albert10} we assume an interstellar absorption 
between 15 and 35 magnitudes and a distance range between 3 and 4 kpc.
In Fig. \ref{sed1} we show the SEDs of the 20 best fitting models. The solid black line represents the best fitting model
from which we obtain the following parameters:
$M_\star\sim23$~M$_{\odot}$, $M_{env}\sim1280$~M$_{\odot}$, and $\dot{M}_{env}\sim0.03$~M$_{\odot}/yr$.
We do not report the disk mass because is not well constrained in the SED fitting. As \citet{robi07} point out, 
in the early stages of evolution, when the disk is deeply embedded inside the infalling
envelope, the relative contributions of the disk and envelope to the SED are difficult to disentangle.

\begin{figure}
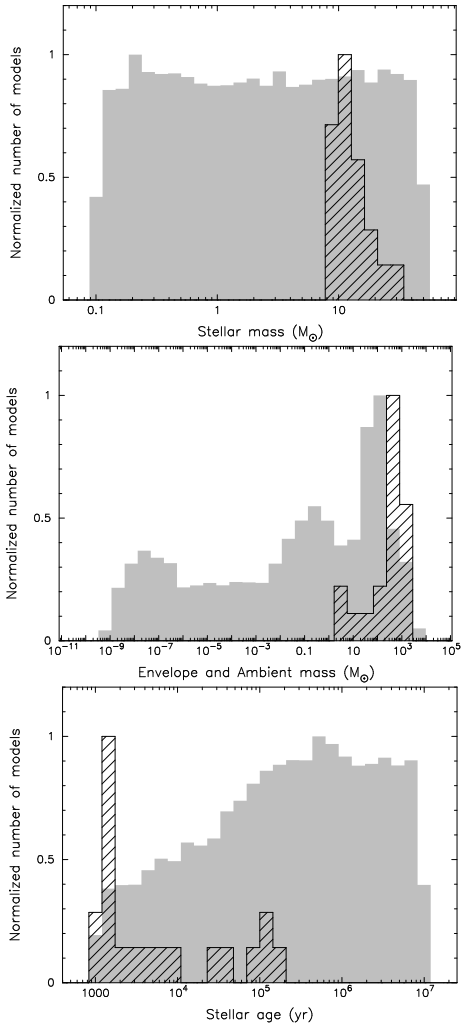

\centering
\includegraphics[height=4.5cm]{hist_mass.ps}
\includegraphics[height=4.5cm]{hist_env.ps}
\includegraphics[height=4.5cm]{hist_age.ps}
\caption{Distribution of some physical parameters (central source mass, envelope mass, and age)
of the 20 best fitting models (hashed columns) together with the distribution of all the models (gray columns).}
\label{sed2}
\end{figure}

Following the \citet{robit06} evolutionary classification, the best fitting model (and also the following 80 fitting
models) correspond to a stage I YSO, i.e., a protostar with large accretion envelope.
In Fig. \ref{sed2} we present the histograms with the distribution of the constrained
physical parameters.
The hashed columns represent the 20 best fitting models and the gray columns correspond to all the models of the grid.
We remark that the 20 best fitting models correspond
to massive central sources (between 8 and 20~M$_{\odot}$) surrounded by massive envelopes
with large accretion rates. Regarding the age of the source, there is a high dispersion in the results
(ages range from $10^{3}$~to $10^{5}$ years) but the presence of a massive envelope is a strong evidence
that the source is at the earliest evolutionary stage.

\subsection{Radio continuum analysis}

\begin{figure}
\centering
\includegraphics[width=7.5cm]{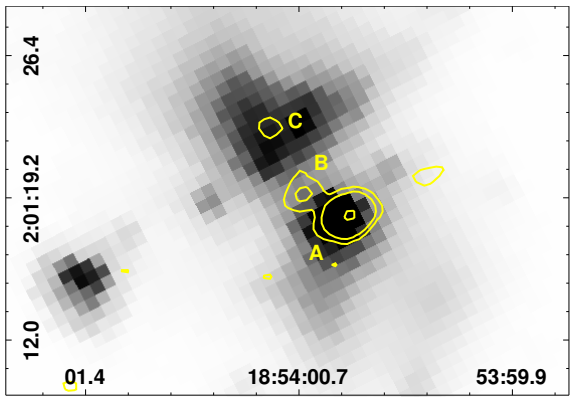}
\caption{Radio continuum emission at 8.4 GHz displayed in contours of 0.3, 1, and 10  mJy beam$^{-1}$ over the 
IRAC-{\it Spitzer} 4.5 $\mu$m emission. The radio continuum has a beam of 1\farcs4 $\times$ 1\farcs0 while the angular resolution of the
mid-IR emission is about 1\farcs5.}
\label{radio}
\end{figure}

We investigate the radio continuum emission at 8.4 GHz of EGOg35.
To produce the image we used archival data from observations
performed with the VLA telescope operating in its B configuration  on
May 7 and 14, 2009 (project code AC948). Data processing was carried out
using MIRIAD software package. The final image has an angular resolution 
of 1\farcs4 $\times$ 1\farcs0 and an rms noise of 0.1 mJy beam$^{-1}$.
Figure \ref{radio} shows, in contours, the radio continuum emission of EGOg35 over its
mid-IR emission at 4.5 $\mu$m. 
As can be seen from this figure, there are three radio sources: source A which coincides with the IR southwestern lobe of the EGOg35, 
source B that appears in between the IR lobes, and source C, marginally detected, which is projected onto the northeastern lobe of the
EGO. The radio source A, the brightest and the only one resolved with this
data set, is centered at 18\hh 54\mm 00.49\ss, $+$02\d 01\m 18.20\s~(J2000), and has an integrated
flux density of 13.5 mJy. The source B is centered at 18\hh 54\mm 00.66\ss, $+$02\d 01\m 19.40\s~(J2000) 
and has a peak intensity of 1 mJy beam$^{-1}$, and source C is centered at
18\hh 54\mm 00.76\ss, $+$02\d 01\m 22.62\s~(J2000) and its peak has a flux density of 0.3 mJy beam$^{-1}$.
Source A and B were also detected at 44 GHz with a similar angular resolution by
\citet{cyga09}. The authors reported an integrated flux density of 12.7
mJy for the brightest radio source and a peak flux density of 3.6 mJy beam$^{-1}$ 
for source B. We calculate the spectral index $\alpha$ (S $\propto \nu^{\alpha}$) between
both frequencies, being  $\alpha \sim -0.05$ for source A, and $\alpha \sim +0.77$ for source B. 
The radio spectral index of the radio source A is consistent with emission from an UCHII region, while the value obtained for source B  
suggests that its radio emission could be due either to an HCHII region \citep{kurtz05b} or to a constant-velocity ionized wind source 
\citep{panagia75}.

From this radio continuum analysis we can suggest the presence of several young stellar objects in the region.
The fact that these radio sources overlap the 4.5~$\mu$m emission reinforces the hypothesis that the
infrared emission originates in YSO shocks.

\section{Summary}

The extended green object EGO G35.03$+$0.35 (EGOg35, in this work), a massive YSO, is embedded in a molecular clump located at the border of an 
HII region. In this work we investigated the 
surrounding molecular gas through several molecular species using the Atacama Submillimeter Telescope Experiment (ASTE). We observed
and analysed the \2 and \3 J=3--2, HCO$^{+}$~J=4--3 and CS J=7--6 transitions, which are useful to trace outflows and dense gas.
To complement our analysis we used IR and radio continuum data from public database. In what follows we summarize the main results of our work:

(1) Most of the molecular spectra observed towards the surveyed region are far of having a simple Gaussian shape, presenting asymmetries, 
absorption dips, and spectral wings or shoulders, characteristics that strongly suggest the existence of kinematical perturbations in 
the gas originated in the object EGOg35.

(2) The \2 J=3--2 line towards EGOg35 shows a double peak structure like with a main component centered at $\sim$54 \k, 
a less intense component centered at $\sim$48 \k, and an absorption dip at $\sim$51.5 \k. The line also presents spectral wings due to the
ouflowing activity of EGOg35. We obtained a total mass and kinetic energy for the YSO outflows of 30 \msol~and 3000 \msol [\k]$^{2}$
($6 \times 10^{46}$ ergs), respectively, similar to those found towards other massive and energetic molecular outflows driven by high-mass YSOs. 

(3) We discovered a HCO$^{+}$ clump towards EGOg35, supporting the presence of outflows in 
the region because it is known that such molecular species enhances in the boundary layer between the outflow jet and the surrounding
molecular core. The central HCO$^{+}$ spectra present an absorption dip at v $\sim54$ \ks due to self-absorbed gas, which 
is probably revealing the existence of a density gradient in the clump. 

(4) We discovered a CS clump towards EGOg35. The CS J=7--6 emission maps the dense envelope where the massive YSO is evolving, and its detection 
implies the presence of molecular gas with temperatures and densities above 40 K and $6 \times 10^{6}$ cm$^{-3}$, respectively.

(5) The central HCO$^{+}$ and CS spectra, which survey the densest region where EGOg35 is embedded, show two velocity components with
the closer one (blue component) stronger than the farthest one (red component). This signature suggests that these lines are mapping 
infalling gas, which is consistent with the presence of a YSO accreting material.

(6) From the spectral energy distribution (SED) study we confirm that EGOg35 is a massive YSO at the earliest 
evolutionary stage (i.e. a class I YSO).

(7) Analysing radio continuum emission towards EGOg35 we conclude that there is evidence of the presence of an 
UCHII region and another source that could be either an HCHII region or a constant-velocity ionized wind source.
The radio continuum analysis suggests the presence of several possible young stellar objects in the region. Future
multiwavelength observations with very high angular resolution will be of a great importance to go deeper in the study of this region.

\section*{Acknowledgments}

We wish to thank the anonymous referee whose comments and suggestions have helped to improve the paper.
S.P., M.O., E.G. and G.D. are members of the {\sl Carrera del
investigador cient\'\i fico} of CONICET, Argentina. A.P. is a doctoral fellow of CONICET, Argentina.
This work was partially supported by Argentina grants awarded by Universidad de Buenos Aires, CONICET and ANPCYT.
M.R wishes to acknowledge support from FONDECYT (CHILE) grant No108033.
She is supported by the Chilean {\sl Center for Astrophysics} FONDAP No.
15010003. S.P. and M.R. are grateful to Dr. Shinya Komugi for the support received during the observations.

\label{lastpage}
\end{document}